# Pulse Generation Based on a Microstrip Circuit with Fourth Order Degenerate Band Edge

Dmitry Oshmarin, Alireza Nikzamir, Michael M. Green, and Filippo Capolino

*Abstract*—A pulse generation scheme is proposed based on a structured resonance in a cavity where the non-conventional energy distribution is concentrated in its middle part. The cavity is first used as an oscillator during the energy charging step and when a switch is activated the signal is extracted from its center. The key component of the proposed scheme is the periodic microstrip waveguide with a fourth-order degenerate band edge (DBE) of its wavenumber-frequency dispersion diagram. The DBE is an exceptional point degeneracy condition that is responsible for the energy to be localized at the cavity center and that can also have the quality factor easily destroyed by a perturbation. The waveguide is designed to have a DBE frequency of 2.86 GHz and produces pulses of approximately 0.1 V peak and 1.1 ns width.

## I. INTRODUCTION

High-power pulse compression/generation is a method in which a high-$Q$ resonant cavity is filled with energy for a duration of time, after which the quality factor is altered by the connection to a load, allowing for the built-up energy to be channeled out to the load [1]–[3]. Alternatively, a high-$Q$ cavity can first be made to oscillate; after the oscillating structure reaches steady-state, the $Q$ of the structure is significantly lowered, allowing the trapped energy of the steady-state oscillation to be channeled out of the structure to an external load [4], [5].

During the 1970s and 80s, great effort was put into the research of increasing the energy of the two-mile-long beam produced by the Stanford Linear Accelerator Center (SLAC), in which an improvement to a pulse compression scheme based on a periodic waveguide was demonstrated [6]–[8]. These studies gave rise to various other schemes for pulse compression and generation devices, including employment of delay lines for radio frequency (RF) applications.

The use of delay lines, in which a propagating electromagnetic wave experiences a considerable reduction of its group velocity $v_g$, is the most common method for RF pulse generation [1], [9], [10]. The simplicity of fabrication, abundance of geometries, and scalability make them an ideal candidate for various types of frequency-selective structures.

The key feature of the scheme described in this paper is the use of a cavity with a structured resonance [11][3], i.e., a resonance associated with the degenerate band edge (DBE), as explained later on. The concept of DBE in periodic structures was pioneered by Figotin and Vitebskiy [11], [12]. In [3], the concept of a pulse compression scheme based on DBE phenomena was provided by using an ideal transmission line formulation, which showed that the energy in a DBE cavity is concentrated at the center of the structure, as shown in Fig. 1(a). The active device and realistic switch were not considered in [3], since that paper focused only on the ideal concept of energy located at the center of the cavity that can be easily extracted when connected to a load.

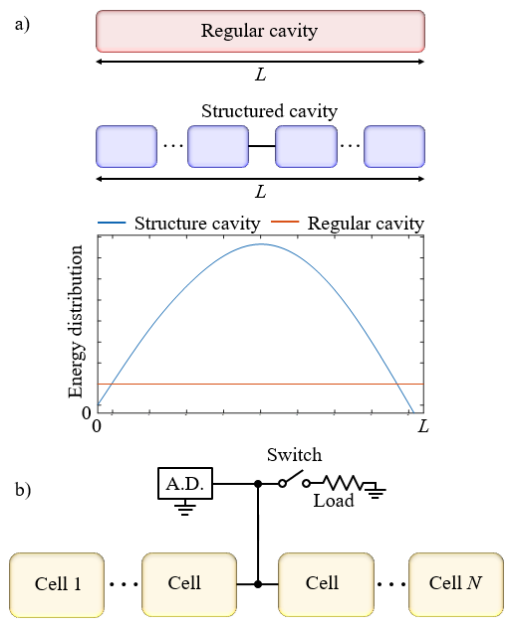

Fig. 1. (a) Schematic of the energy distributions inside a regular cavity and inside a structured DBE resonant cavity; for the structured cavity, the non-conventional energy distribution is mainly concentrated at the physical center. (b) Schematic representation of the pulse generation scheme based on a structured cavity able to support DBE resonance. The negative resistance active device (A.D.) is inciting oscillations in the cavity made of $N$ unit cells during the "charging time". Then, the switch to a load is used to drastically alter the loaded quality factor, $Q$, of the structure and extract the energy.

This paper focuses on a realistic design of a DBE cavity and the experimental demonstration of the pulse generation scheme, including (i) the charging time when the cavity is made unstable and starts oscillations, and (ii) the extraction time when the load is connected to the center of the cavity via a switch (Fig. 1 (b)). The presented material is base on what was previously presented in [3], [13] that is not repeated here for brevity. We focus on a particular type of high-Q cavity made of two coupled microstrip waveguides, each consisting of $N$ cascaded unit cells forming a periodic structure. Wave propagation in the waveguide is described by the evolution of eigenmodes from one cell to the next. When these eigenmodes coalesce into a degenerate one, a degeneracy in Bloch-wavenumber-frequency space is also visible as a flat dispersion curve [14], [15]. The number of eigenmodes that coalesce determines the order of the formed exceptional point of degeneracy. In this paper, we focus on a fourth-order degeneracy, known as DBE, that exists in a lossless, periodic structure [12], [14], [16]–[18]. Near the DBE frequency $\omega_d$, the dispersion relation can be approximated as

$$\omega_d - \omega \propto (k - k_d)^4$$

where $k$ is the Floquet-Bloch wavenumber, $k_d$ is the degenerate wavenumber at the edge of the Brillouin zone, and $\omega$ is the angular frequency; the exponent of 4 indicates a fourth-order degeneracy. In this paper, the DBE cavity is not filled by an external signal, but it is rather made to oscillate via an active component located at the center of the cavity (Fig. 1 (b)). The power is extracted by closing a switch and connecting the structure to a load.

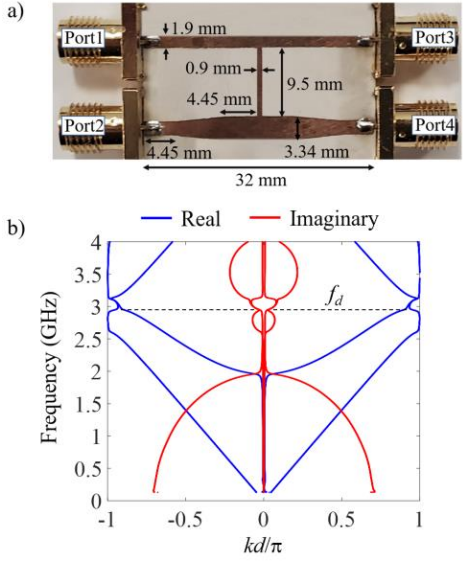

Fig. 2. (a) 4-port unit cell of a periodic structure with DBE used in the pulse generation scheme; (b) Real and imaginary parts of the wavenumber dispersion relation for an infinitely long periodic structure made up of unit cells in (a), showing the occurrence of the DBE at 2.93 GHz and denoted as $f_d$.

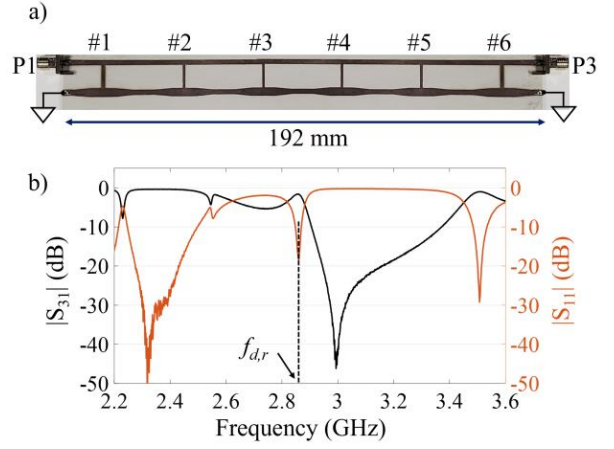

Fig. 3. (a) Two-port, 6-cell structure with DBE used in the pulse generation scheme. Transmission and reflection characteristics are measured from Port1 (P1) to Port3 (P3) to determine the location of the DBE resonance $f_{d,r}$. The lower microstrip is short circuited at its left and right terminations; (b) Transmission and reflection coefficients for the 6-cell structure showing the DBE resonance at 2.86 GHz, denoted by the dip of $|S_{11}|$ and almost unity $|S_{31}|$. At higher frequencies, up to 3.4 GHz, attenuation in transmission is due to the bandgap.

The paper is organized as follows. In Sec. II, we discuss how the energy distribution differs between the regular cavity and the structured cavity. Then, we describe the unit cell that forms the basis of the periodic structure, and its wavenumber dispersion relationship. We also show how the number of unit cells to be used in the cavity is determined using Keysight ADS Method of Moments simulator. After obtaining the suitable number of unit cells, $N = 6$ for our design, transmission and reflection characteristics (S-parameters) are measured and discussed. In Sec. III, the active circuit to obtain negative conductance, switch configuration to have ON and OFF states to extract the power to a load, and pulse generating structure are presented, with all of the nuances of the scheme setup and measured results.

## II. PERIODIC STRUCTURE WITH DBE RESONANCE FOR PULSE GENERATION

The schematic for the proposed pulse generation architecture is shown in Fig. 1(b). The sum of electric and magnetic energies is concentrated at the center of the structure operating at the DBE resonance [3], in contrast to what happens in a cavity made of a truncated, standard waveguide, where the energy is uniformly distributed along the cavity as shown in Fig. 1(a) with the solid red line. For the periodic DBE cavity with length $L$, the energy distribution curve has a maximum at the middle of the structure, as shown in Fig. 1(a) with the solid blue line. The results were obtained by the use of Keysight ADS in which a 1 V AC voltage source was connected to the left of both structures, and the energy was sampled (at the resonance frequency of interest, which is discussed later) along the length of both structures, and then plotted versus length.

For the pulse generation scheme, first, the cavity is filled with energy, as described next, by establishing oscillations with a frequency slightly below the DBE frequency, where the DBE resonance of the DBE cavity is located, and then an external switch attached to the periodic structure is closed to extract the trapped energy.

As described in Sec. I, a periodic structure consisting of $N$ unit cells utilizes an active device (A.D.) configured to produce negative resistance through a positive feedback mechanism. The periodic structure itself acts as a frequency selective cavity, where the most important resonance (called DBE resonance) is the one closest to the DBE frequency [11], [13]. Once the structure is driven to start oscillations and reaches steady-state, a switch connecting the center of the periodic structure shorts it to an external load. The connection of the load drastically alters the loaded quality factor, $Q$, allowing for the trapped energy to be channeled to the load producing a narrow-in-time pulse. Due to the unique property of an oscillator-based structure with DBE, which is studied in [19], [20], the time required to reach a steady state can be considerably longer than for a simple LC oscillator or a comparable single-ladder oscillator [19]. Thus, if the highest possible amplitude of the extracted pulse is required, the time periodicity of the generated pulses is much longer than the generate pulse duration. On the other hand, if the shortest time between extracted pulses is required, it is unnecessary to wait for the structure to reach steady-state, and in this case, the switching properties of the switch, like rise and fall times, become the limiting factors. In our case, the characteristics of the switch are the most crucial parameters in designing a narrow-in-time pulse with high amplitude. The switch needs to alter the quality factor as quickly as possible.

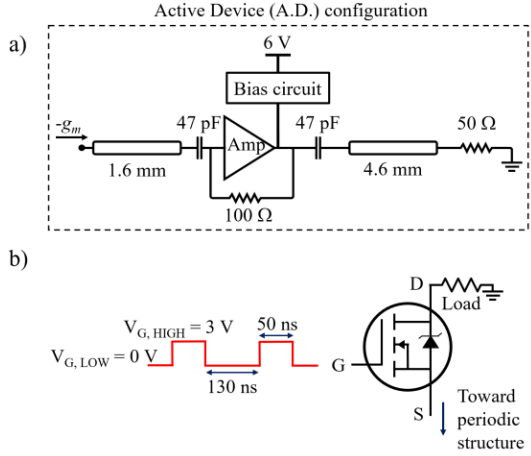

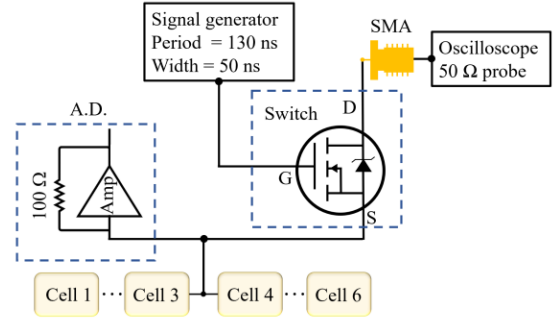

Fig. 4. (a) Schematic representation of the BGA6589 active device (A.D.) with a 100 Ω feedback resistor able to produce -2 mS of negative conductance. The device is configured and biased as suggested by the manufacturer at the frequency of operation; (b) IRFZ44N switch schematic that is controlled by a periodic voltage source applied to the gate. The Source (S) is connected to the center of the upper microstrip line of the periodic structure, while the Drain (D) is connected to the load (see Figs. 1(b) and 5). In the actual fabricated and measured structure, an oscilloscope is connected to the Drain (D) of the switch through a probe with an impedance of 50 Ω that acts as the load. The gate voltage is zero while the structure is reaching the steady-state oscillation to avoid loading of the Q.

Fig. 5. Measurement setup functional diagram. First, an active device (A.D.) connected to the center of the 6-cell structure makes the structure oscillate. The signal generator is connected to the Gate (G) of the switch to short the Drain (D) to the Source (S) by applying enough voltage at the Gate (G) to overcome the overdrive voltage of the switch. The energy that was "trapped" inside the 6-cell structure, flows to the 50 Ω probe of an oscilloscope, where a pulse followed by an exponential signal decay is observed.

A fabricated unit cell of the periodic structure in Fig. 1(b), along with its dimensions, is shown in Fig. 2(a). The unit cell is fabricated on Rogers RO3003 substrate with 0.76 mm (30 mils) thickness, dielectric constant of 3, dissipation factor of tan δ = 0.001. The top line (when uncoupled) is designed to have 50 Ω characteristic impedance. The dispersion diagram describing the modes in a waveguide made of an infinite cascade of unit cells in Fig. 2(a) is shown in Fig. 2(b). This diagram was obtained by measuring a 4×4 S-parameter matrix with a Keysight Vector Network Analyzer (VNA), converting it to a transfer matrix with the help of Mathworks MATLAB numerical computing environment, and then evaluating the eigenvalues derived from the transfer matrix at each frequency and converting them to Bloch wavenumbers. See [13], [21] for an in-depth explanation.

From Fig. 2(b), it can be seen that the DBE frequency, $f_d$, is at 2.93 GHz. At a slightly higher frequency, and until approximately 3.2 GHz, the imaginary parts of four eigenmodes, shown in red, are non-zero, representing four evanescent waves, forming a stop band. Slightly below the DBE frequency, two modes propagate and two are evanescent. At the DBE frequency, the four modes coalesce ideally to a real-valued wavenumber. The presence of losses due to imperfections and material losses, which are further exacerbated by an increase in the number of unit cells added to the peridic structure, perturb the DBE as already described in previous publications; e.g., in [13].

Next, we consider a cavity consisting of a periodic structure of finite length, made up of six unit cells in Fig. 2(a). The cavity resonates at the so called DBE frequency, denoted as $f_{d,r}$, very close to the DBE frequency, as shown in [11], [13], [18], [22]. This is also the frequency at which the energy was sampled to obtain the energy distribution curve of Fig. 1(a). Ideally, as the number of unit cells $N$ approaches infinity, the DBE resonance frequency $f_{d,r}$ approaches $f_d$ [11], [15]. In an ideal lossless structure with DBE, the quality factor of the resonator increases with the cavity length as $Q \propto N^5$. However, the DBE is sensitive to losses [13], [18], hence the number of unit cells must be carefully considered. In practice, due to material and radiative losses (among others), each additional cell would contribute to suppress the DBE resonance, and after a particular $N$, the DBE resonance would not show off in the transfer function of the whole cavity. But even if the periodic structure does not contain many unit cells, the DBE resonance can still exhibit some of the desired features of the ideal DBE such as a high-quality factor and energy localization [13], [18]. By utilizing the Keysight ADS Method of Moments simulator, it was determined that 6 cells offer a strong DBE resonance at a frequency slightly below the DBE frequency $f_d = 2.93$ GHz. It is unnecessary to operate at the exact $f_d$ since most of the beneficial properties such as enhanced quality factor can be seen at DBE resonance frequency $f_{d,r}$. The high quality factor (based on the transmission characteristics shown later) is the desirable feature for our application, especially because the DBE's high sensitivity to perturbations allows for this quality factor to be drastically lowered by adding a load at the center of the structure, where most of the energy resides. Once the DBE is perturbed by connecting the load, the energy that was trapped mainly near the center of the structure [3] is quickly channeled to an external load via the created path.

The fabricated 6-cell periodic structure is shown in Fig. 3(a). Two SMA connectors were added at the edges of the upper line to experimentally confirm the existence of the DBE resonance $f_{d,r}$, while the edges of the lower line were shorted to the ground. By measuring the transmission coefficient from Port1 to Port3 ($S_{31}$), and the reflection coefficient of Port1 ($S_{11}$), we are able to confirm the existence of a resonance associated with the DBE condition. At the frequency of 2.86 GHz, there is a dip in $|S_{11}|$ while $|S_{31}|$ is nearly 0 dB. At slightly higher frequencies, $|S_{31}|$ has a sharp drop and $|S_{11}|$ is nearly zero dB, because of the bandgap, as shown in the dispersion diagram, confirming the DBE resonance.

### III. ACTIVE PULSE GENERATION STRUCTURE

In this section, we discuss the active structure realization. The negative resistance/conductance active device is based on the BGA6589 MMIC wideband medium power amplifier by NXP Semiconductors configured in a feedback configuration as shown in Fig. 4(a). The bias circuit, as well as the 47 pF DC blocking capacitors, are chosen and configured as specified by the manufacturer at the desired frequency of operation. The 100 Ω feedback resistor provides $-2$ mS of negative conductance with negligible susceptance at the point labeled $-g_m$ in Fig. 4(a), as explained in [22] with more details. Based on the study done in [22], this conductance is sufficient for the 6-cell structure to sustain oscillation. The switch that connects the load to the periodic structure is the Infineon IRFZ44N Power MOSFET switch, whose configuration is shown in Fig. 4(b). The Source (S) and the Drain (D) of the MOSFET switch are shorted when 3 V is applied to the Gate (G), labeled as $V_G$ in Fig. 4(b). The time span between the high states, i.e., $V_G = 3$ V, of the square wave applied to the Gate (G) will be one of limiting factors in reaching the steady-state and must be paid close attention during the design. In general, this time span should be larger than the time it takes for the structure to reach steady-state.

The functional diagram of the measurement setup is shown in Fig. 5. The middle of the upper line of the 6-cell structure is connected to both the active device circuit and the Source (S) of the switch. While the 6-cell structure is filled with energy in the form of the steady-state oscillation, the Gate (G) of the transistor is biased at 0 V so that the switch is in the "OFF" state; i.e., the Source (S) and Drain (D) of the switch are disconnected. During this state, the 6-cell periodic waveguide ideally does not experience any external loading. Once steady-state oscillation is reached, an external signal generator biases the Gate voltage of the switch to 3 V to short the Drain and the Source. The Drain of the switch is connected to an oscilloscope, with probe impedance set to 50 Ω. In other words, we use the oscilloscope probe's input impedance as the structure's load in our measurement setup. Therefore, during the switch's "ON" state, the load is connected to the waveguide, drastically lowering the quality factor and allowing a way for the energy to get out of the cavity. The oscilloscope is also used to monitor the produced pulse in time, verifying the proper operation of the scheme. The pulse seen on the oscilloscope is an exponentially decaying sinusoidal function still oscillating at a frequency of $f_{d,r}$, the envelope of which should be related to the value of the load connected to the waveguide (50 Ω impedance of the oscilloscope probe) and parasitic capacitances.

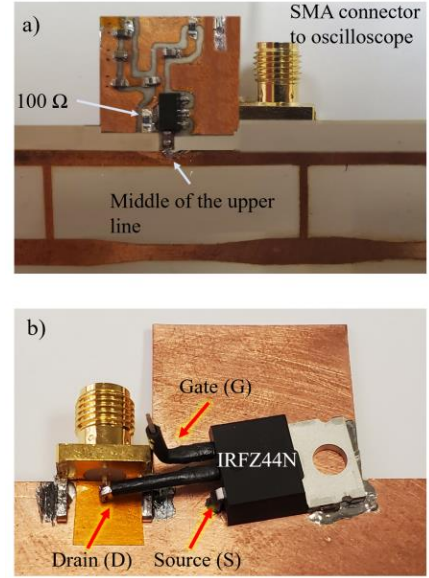

Fig. 6. (a) Active device circuit attached to the middle of the upper line. 100 Ω feedback resistor provides -2 mS of negative conductance. An SMA connector connects the middle of the 6-cell structure, through a switch, to the oscilloscope. SMA connected to the oscilloscope with probe's impedance set to 50 Ω disturbs the DBE resonance and resulting pulses are observed on the oscilloscope; (b) Backside of the structure right behind the active device showing the switch placement and its connection to the SMA pin. Source (S) goes through the substrate to the other side of the microstrip and connects to the middle of the 6-cell structure. All of the switch's leads are wrapped in shrink tubes for isolation. All of the exposed copper is ground.

The practical realization of the Fig. 5 setup is shown in Fig. 6(a) and (b). Fig. 6(a) shows how the middle of the fabricated 6-cell structure is connected to the active device circuit, and Fig. 6(b) shows how the structure is connected to the switch. Just as in Fig. 5, the Drain of the switch is attached to an oscilloscope probe (through an SMA connector) with 50 Ω input impedance, while the Source is connected to the middle of the structure.

First, with the switch in the "OFF" state, we verify that the structure is oscillating at the expected frequency of $f_{d,r}$ by connecting an oscilloscope to Port3, after which both Port1 and 3 are short circuited to the ground. (Note that a DBE cavity is not very sensitive to any load connection at the edge of the cavity [21].

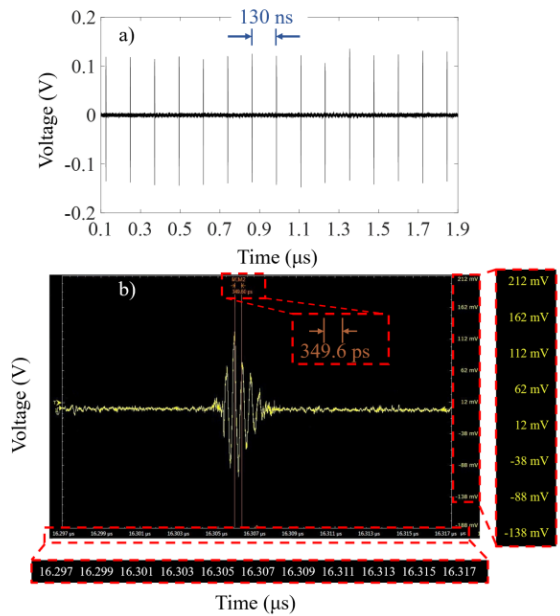

Fig. 7. (a) Time domain pulse seen on the oscilloscope when the switch is turned ON every 130 ns and stays open for 50 ns; (b) Captured image of the zoomed-in pulse shape as is seen on the oscilloscope. The period of the oscillation within the envelope is exactly $1/f_{d,r}$ or 349.6 ps confirming that prior to the switch turning ON, structure was oscillating at DBE resonance frequency $f_{d,r}$. The top inset showing the zoomed in marker measurement between two adjacent peaks. Two axis insets are added for clarity. The pulse length is approximately 3 period long.

Fig. 7(a) shows the extracted pulse train that is produced when the switch closes for 50 ns ("ON" state) and stays in the "OFF" state for 130 ns. Fig. 7(b) shows a close-up of one of the pulses showing that all of the energy is extracted in approximately three oscillations, i.e., around 1.15 ns, and the period of oscillation is 349.6 ps or $1/f_{d,r}$. The remarkable thing to note here is that while it takes more than 100 ns for the structure to reach steady-state oscillation, it takes less than 2 ns to extract most of the energy in the form of a pulse [22]. This is due to the sensitivity of the DBE to perturbations, such as the loading at the center when the switch is in the "ON" state.

## IV. CONCLUSION

A pulse generation scheme has been proposed based on a structured resonance with fourth-order exceptional point degeneracy, called DBE, producing pulses having approximate width of 1.1 ns. The key feature of this new scheme is the structured resonance where the unconventional energy distribution concentrates in the middle of the cavity, instead of being uniformly distributed in the whole cavity length as in conventional cavities. The DBE is the enabling condition to obtain the structured resonance. The scheme is very versatile in its design, and key performance characteristics can be altered by a simple geometry or component change without major alterations to the conceptual design. The performance characteristics of the presented pulse generation scheme can be altered, e.g., the pulse width could be controlled by a proper selection of the load. The period of the produced pulses can be changed by the signal generator that controls the Gate of the switch. If higher power of the extracted pulse is desired, the DBE cavity should be redesigned to contain higher voltage levels at its center, though this is a subject that requires more investigation because still many features of the DBE resonance are not well understood. The experimental results in this paper constitute the first ever demonstration that a DBE structured cavity can be used for pulse generation, showing a frequency of oscillation corresponding to the DBE resonance. Several aspects could be improved in the future including the location of the switch and active component, and by comparing the DBE pulse generator with other conventional devices using a Fabry-Perot cavity. The results also guide the direction for future developments of this pulse generation scheme. Longer cavity lengths could be used if material losses are small (like in a fully metallic structure) or if distributed gain is considered as in the distributed DBE oscillator in [20].

The same scheme can be adopted at very high power where the DBE cavity shall be made of coupled modes in a metallic hollowed waveguide where the quality factor is altered via a plasma switch [3], [23]–[27].

## V. ACKNOWLEDGMENT

This material is based upon work supported by the National Science Foundation under award NSF ECCS-1711975.